\documentclass{PoS}

\usepackage[T1]{fontenc}
\usepackage{verbatim}
\usepackage{mathtools}
\usepackage{physics}
\usepackage{slashed}
\usepackage{rotating}
\usepackage{url}
\usepackage[toc]{appendix}
\usepackage{braket}

\def\cbar{\overline{c}}

\def\ellbar{{\overline{\ell}}}

\def\Heff{\mathcal{H}_{\rm eff}}

\newcommand{\bea}{\begin{eqnarray}}
\newcommand{\eea}{\end{eqnarray}}
\newcommand{\beq}{\begin{equation}}
\newcommand{\eeq}{\end{equation}}
\newcommand{\ec}{\end{center}}
\newcommand{\bc}{\begin{center}}

\title{Study of Anomalies in Exclusive Semileptonic B Decays }

\ShortTitle{Anomalies in B decays}

\author{\speaker{Vincenzo Afferrante}\footnote{Supported by the FWF DK W1203-N16}\\
       University of Graz\\
        E-mail: \email{vincenzo.afferrante@uni-graz.at}}

\author{Guido Martinelli\\
        Sapienza University of Rome\\
        E-mail: \email{guido.martinelli@roma1.infn.it}}
    
\abstract{
	We discuss the anomalies observed in the semileptonic decays of the $B$ into the $D$ and $D^*$ mesons, and the difference between the exclusive and the inclusive determination of the CKM matrix element $V_{cb}$. Our analysis is based on Belle data with unfolded kinematical dependence.  We use two different parameterizations of the form factors, the CLN and BGL parameterizations. They mainly differ on  the form factors which can be determined within the HQET, with CLN that relies more on the latter. It is notable that the two parameterizations give different results for $V_{cb}\,$. 
	
	We investigate the ratio of the decay $B \to D^* \tau \nu$ compared to the decays into light leptons. The experimental value of the ratio is systematically larger than the Standard Model (SM) prediction. Extensions of SM with a scalar or pseudo-scalar effective operator may give an enhancement of the decay rate of the B meson into the lepton of the third generation. The effects of such new physics operators about these decays are studied.}

\FullConference{An Alpine LHC Physics Summit (ALPS2018)\\
		15-20 April, 2018\\
		Obergurgl, Austria}

\begin{document}

\section{Belle Data Analysis}

\begin{figure}
	\centering
	\includegraphics[width=0.5\textwidth]{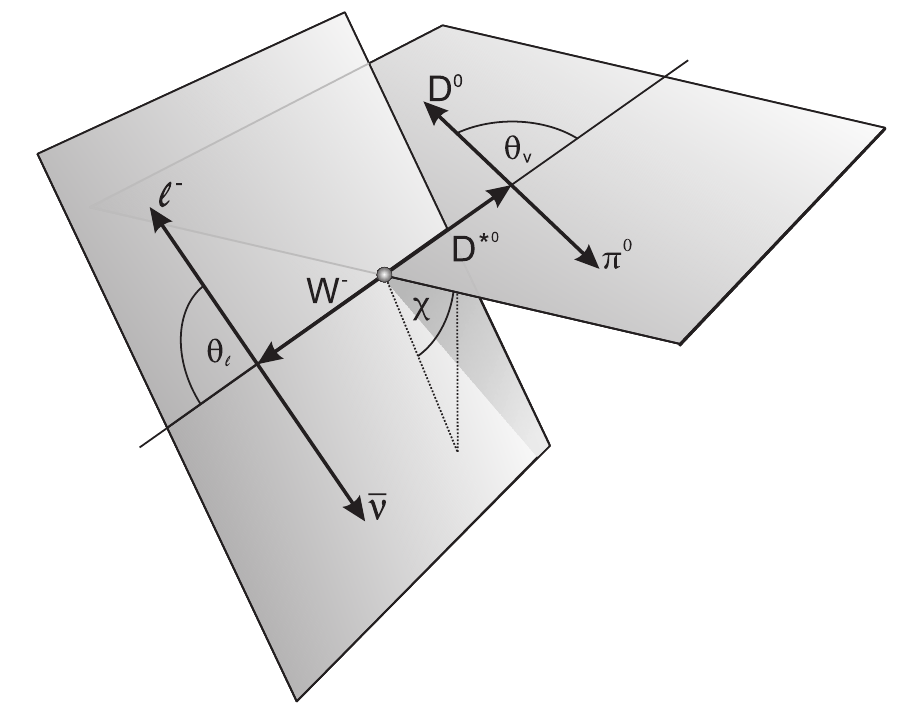}
	\caption{The helicity angles used for the study of helicity amplitudes $\theta_l$, $\theta_v$ and $\chi$ as shown in \cite{abdesselam2017precise}. The former two of these angles are defined in the center of mass of the decay products,  the virtual $W^-$ and $D^*$. The angle $\theta_l$ is defined by the directions of the charged lepton and the direction of $W$ in the $W$ rest frame. Similarly $\theta_v$ is defined using the directions of the $D^*$ and the $D$ in the $D^*$ rest frame. The angle $\chi$ is obtained between the two planes where the $\theta_l$ and $\theta_v$  are defined.}
	\label{fig:angles}
\end{figure}

We  use the data from a recent Belle analysis \cite{abdesselam2017precise} that provides the unfolded full differential decay rates of $B \rightarrow D^* l \nu$, described in terms of the recoil variable $w$, which can be defined in terms of the invariant mass squared of the leptonic couple \begin{equation}
w = \dfrac{	m_B^2 + m_{D^*}^2 - q^2}{2 m_B m_{D^*}}
\end{equation} and of the three kinematical angles depicted in figure \ref{fig:angles}. Since the only input that comes from the lattice is the axial form factor at zero recoil, to obtain the full dependence over the recoil range it is necessary to parametrize the form factors using constraints that come from unitarity, crossing symmetry, quark-hadron duality and HQET. These constraints are subject to some systematical uncertainties, which are difficult to estimate.  We now give details about the parameterizations which we used, we explain the methods used of the fit, then we give our results.

One of the parametrizations of the hadronic form factors, was given by Caprini, Lellouch and Neubert \cite{caprini1998dispersive}.  The other is the one by Boyd, Grinstein and Lebed \cite{boyd1997precision}. 
The two differ principally on assumptions on the validity of HQET at the first order. 
We will discuss  how to use these information about the factors, in order to analyze the existing experimental data.

\section{The BGL Parametrization}

We describe now how we used the BGL parameterization for the analysis. The decay formula in terms of the variable $w$, the three kinematic angles, described in figure \ref{fig:angles}, and the helicity amplitudes is given by

 \begin{equation}
\begin{split}
\dfrac{d \Gamma(B \rightarrow D^*(\rightarrow D \pi) l \nu)}{d w d \cos \theta_v d \cos \theta_l d\chi} = & \dfrac{  6 \eta_{EW}^2 m_B m_{D^*}^2 G_F^2 |V_{cb}|^2 \sqrt{w^2 -1} (1 + r^2 - 2 w r) }{8 (4 \pi)^4 m_B^2} B(D^* \rightarrow D \pi) \times \\ 
& \{   (1 - \cos \theta_l )^2 \sin^2 \theta_v | H_{++}|^2 \\
&+ (1 + \cos \theta_l )^2 \sin^2 \theta_v |H_{--}|^2 
+ 4  \sin ^2 \theta_l \cos^2 \theta_v |H_{00}|^2   \\
& - 2 \sin^2 \theta_l \sin^2 \theta_v \cos 2 \chi H_{++} H_{--} \\ 
&- 4 \sin \theta_l (1 - \cos \theta_l) \sin \theta_v \cos \theta_v \cos \chi H_{++} H_{00} \\ 
& +4 \sin \theta_l (1 + \cos \theta_l)\sin \theta_v \cos \theta_v \cos \chi   H_{--} H_{00} \} 
\end{split}
\end{equation}
 and the matrix elements are written in terms of form factors as in ref. \cite{boyd1997precision} \begin{align}
\braket{D^*|V_\mu|B} & = i g(w) \epsilon_{\mu \alpha \beta \gamma} \epsilon^\alpha p_B^\beta p_{D^*}^\gamma \,,  \\
\braket{D^*|A_\mu|B} & = f(w) \epsilon_\mu + (\epsilon \cdot p_B)[a_+(w) (p_B+p_{D^*})_\mu + a_-(w) (p_B - p_{D^*})_\mu ] \,.
\end{align} The helicity amplitudes are expressed in terms of the factors as \begin{align}
H_{00}(w) = \dfrac{\mathcal{F}_1(w)}{\sqrt{q^2}} \; \; ; \; \;
H_{\pm \pm } = f(w) \mp m_B m_{D^*} \sqrt{w^2 -1} g (w) \,,
\end{align} where the factor $\mathcal F_1$ is defined by \begin{equation}
\mathcal F_1(w ) = \dfrac{1}{m_{D^*}}[2 m_B^2 m_{D^*}^2(w^2-1)a_+(w) + m_B m_{D^*} w f(w) ] \,.
\end{equation} We write the expansion for the three form factors  as \begin{align}
g(z) = \dfrac{1}{P_g(z)\phi_g(z)} \sum_{n=0}^{N} a_n z^n \; \; ; \; \;
f(z) = \dfrac{1}{P_f(z)\phi_f(z)} \sum_{n=0}^{N} b_n z^n \; \; ; \; \; \mathcal{F}_1(z) = \dfrac{1}{P_{\mathcal{F}_1}(z)\phi_{\mathcal{F}_1}(z)} \sum_{n=0}^{N} c_n z^n \,.
\end{align} By unitarity, we may impose bounds on the coefficients of the expansions $\{a_n, b_n, c_n\}$. The coefficients have to respect the following conditions \begin{equation}
\sum_{n=0}^N |a_n|^2 \leq 1 \;\;;\;\; \sum_{n=0}^N (|b_n|^2 +|c_n|^2) \leq 1 \,.
\end{equation} The outer functions $\phi$ are phase space factors, while the inner functions $P$ are made of products of Blaschke factors, which then remove poles associated to the production of excited $B_c$ meson states, with mass smaller than $m_B + m_{D^*}$.

\section{The CLN Parametrization}

 We write the relevant (nonzero) helicity amplitudes $H_{m m'}$, in terms of the form factors used in the work of Caprini et al.  \begin{equation}
\begin{split}
H_{\pm \pm} (q^2) &= ( m_B + m_{D^*}) A_1(q^2) \mp \dfrac{2 m_B}{m_B + m_{D^*}} |\mathbf{q}| V(q^2) \,, \\
H_{00} (q^2) &= \dfrac{1}{2 m_{D^*} \sqrt{q^2} } \left[(m_B^2 - m_{D^*}^2 -q^2)(m_B + m_{D^*})A_1(q^2) - \dfrac{4 m_B^2 |\mathbf{q}|^2 }{m_B + m_{D^*}} A_2(q^2)  \right] \,, \\ 
H_{0t} (q^2) &= \dfrac{2 m_B |\mathbf{q}|}{\sqrt{q^2}} A_0(q^2) \,.
\end{split}
\label{helicity_amplitudes}
\end{equation}
In order to exploit the Isgur-Wise limit and its corrections, the four form factors in the helicity amplitudes \eqref{helicity_amplitudes} are expressed in terms of a universal form factor $h_{A_1}(w)$ and three ratios $R_i(w)$, defined as \begin{align}
A_1 = \dfrac{w+1}{2} r' h_{A_1}(w) \; \; ; \; \; A_0 = \dfrac{R_0(w)}{r'} h_{A_1}(w) \; \; ; \; \;
A_2 = \dfrac{R_2(w)}{r'} h_{A_1}(w) \; \; ; \; \; V= \dfrac{R_1(w)}{r'} h_{A_1}(w)
\label{eq:form_factors_ratio}
\end{align} where \begin{equation}
r' = \dfrac{2 \sqrt{m_B m_{D^*}}}{m_B + m_{D^*}} \,.
\end{equation} The $w$ dependence of the universal form factor and the ratios are heavily constrained by the HQET and have the following form, in terms of five parameters:  $h_{A_1}(1)$, which is the universal form factor at  minimum recoil,  $R_i(1)$ that are the ratios at minimum recoil, and $\rho^2_{D^*}$, which is the only parameter governing the slope of the form factors \begin{align}
h_{A_1}(w) &= h_{A_1}(1)(1 - 8 \rho^2_{D^*} z + (53 \rho^2_{D^*} - 15 )z^2 - (231 \rho^2_{D^*} -91 ) z^3 ) \,, \\
R_0(w) &= R_0(1) -0.11(w-1) + 0.01(w-1)^2 \,, \\
R_1(w) &= R_1(1) -0.12(w-1) + 0.05(w-1)^2 \,,\\
R_2(w) &= R_2(1) + 0.11(w-1) -0.06(w-1)^2 \,.
\label{eq:CLN_form_factor}
\end{align}  With massless leptons the ratio $R_0$ is not used.

\section{Fit Methods and Results}

The two parameterizations we just described are used to fit the unfolded spectrum given by the Belle collaboration, as done in \cite{bigi2017fresh} and \cite{grinstein2017model}. The fit was realized with  the software BAT (Bayesian Analysis Tools) \cite{caldwell2009bat}, which use Markov Chain Monte Carlo to obtain parameters, with their uncertainties. The CLN parametrization depends on four parameter to fit, namely $V_{cb} h_{A_1}(1)$, $ \rho^2_{D^*}$, $R_1(1)$ and $R_2(1)$, while the BGL parametrization depends on six parameters, which are $V_{cb}$, $a_0, a_1, b_1, c_1, c_2$. We have decided to truncate the series at second order since  in the semileptonic range $z_{max} \approx 0.056$. It is notable that CLN has only one slope parameter, $\rho^2_{D^*}$, while BGL has more parameters that control the functional dependence over kinematical range, outside the minimal recoil region. 

The unfolded Belle data are presented in 4 histograms, each of ten bins, and with a correlation matrix between the 40 bins. The histograms show $d \Gamma/ dx$, where $x$ is one of the four kinematical variables described before: $w, \cos \theta_v, \cos \theta_l $ and $\chi$.

\begin{figure}
	\centering
	\includegraphics[width=\textwidth]{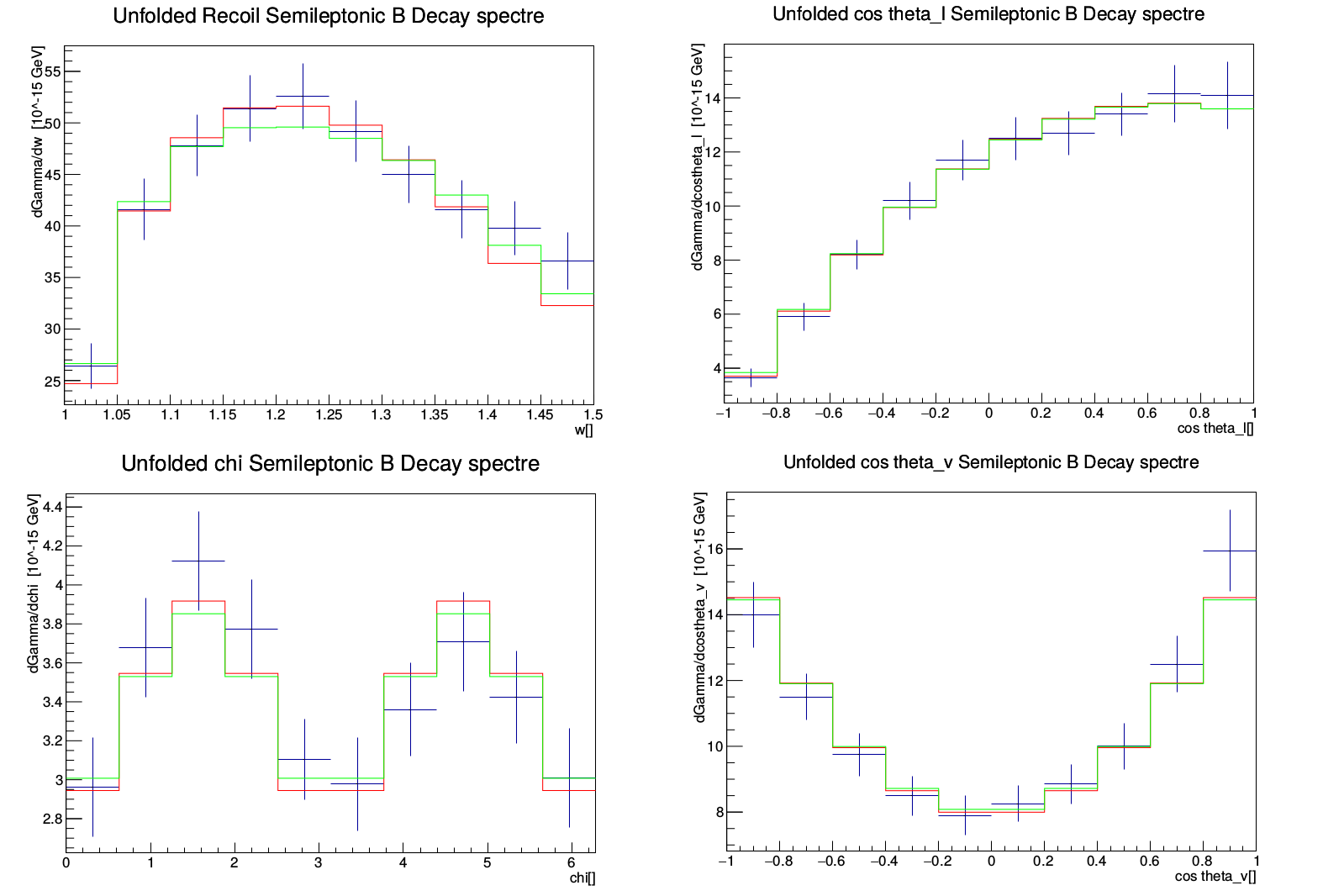}
	\caption{Histograms that confronts the data (in blue) for the unfolded spectrum in terms of the four variables, with the result of the fit for CLN (in red) and BGL (in green). }
	\label{fig:histograms}
\end{figure}

We proceeded to fit all the Belle data, using the numerical integration of the full differential formula in every one of the 40 bin.
We show the histograms of the data with their error, together with the fit obtained using both the CLN and the BGL parametrizations, in the figure \ref{fig:histograms}. The results obtained with the CLN parametrization are given in table \ref{tab:CLN_results}.

\begin{table}
	\centering
	\begin{tabular}{ |c|c | }
		\hline
		CLN Parameters & Mean Values $\pm$ Errors \\
		\hline 
		$V_{cb} h_{A_1}(1) $ & (35.662 $\pm$ 1.132) $10^{-3}$ \\
		$R_1(1)$ & 1.474 $\pm$ 0.072  \\
		$R_2(1)$ & 0.9105 $\pm$ 0.0723 \\
		$\rho_{D^*}^2$ & 1.196 $\pm$ 0.140 \\
		\hline
	\end{tabular}
	\caption{Results from the fit with CLN parametrization are shown}
	\label{tab:CLN_results}
\end{table}

 In the fits we used value of the axial form factor at zero recoil taken from \cite{bailey2014update} \begin{equation}
 \label{lattice_form_factor}
 h_{A_1}(1) = 0.906 \pm 0.013 \;.
 \end{equation} 
In the case of BGL $h_{A_1}(1)$ fix one of the parameter of the expansion, namely $b_0$, to the value \begin{equation}
b_0 = 2 P_f(0) \phi_f(0) \sqrt{m_B m_{D^*} } h_{A_1}(1) = 0.01223 \pm 0.00018 \,.
\end{equation} The results for BGL are shown in the table \ref{tab:BGL_result}.
\begin{table}
	\centering
	\begin{tabular}{ |c|c | }
		\hline
		BGL Parameters & Mean Values $\pm$ Errors \\
		\hline 
		$V_{cb}  $ & (42.1 $\pm$ 1.2) $10^{-3}$ \\
		$a_0$ & 0.0126 $\pm$ 0.0056  \\
		$a_1$ & 0.71 $\pm$ 0.19 \\
		$b_1$ & -0.0428 $\pm$ 0.020 \\
		$c_1$ & -0.0071 $\pm$ 0.0053 \\
		$c_2$ & 0.066 $\pm$ 0.094 \\
		\hline
	\end{tabular}
	\caption{Results from the fit with BGL parametrization are shown}
	\label{tab:BGL_result}
\end{table}

It's remarkable that the result for $V_{cb}$ obtained with BGL parametrization is sensibly higher than that obtained with CLN parametrization. The BGL results are also much closer to the result that comes from the inclusive determination of $V_{cb}$, precisely
\begin{equation}
|V_{cb}| = (42.00 \pm 0.65) 10^{-3} \,.
\end{equation}  This last result is obtained using Heavy Quark Expansion and fitting various kinematical variables from inclusive decay data, as explained in \cite{gambino2016taming}.

It is safe to say that, without more information from the lattice, a parametrization that does not use HQET relations, but it is based principally on unitarity bounds gives a more reliable extrapolation of $|V_{cb}|$ from exclusive decays. 

\section{Search of New Physics Effects in Belle Data}

We want to study the presence of effects of NP at low energies, from the Belle data used for the extrapolation of $V_{cb}$. The most general way to include them in the analysis is by modifying the effective Hamiltonian, adding non-SM terms that respect Lorentz invariance. We keep a $V-A$ form for the leptonic part and we parametrize the couplings of new operators with five coefficients $g_i$, $i = V, A, S, P, T, T_5$ that multiplies operator densities not present in the SM. Obviously, putting $g_i =0$ will give the SM result, with the $V-A$ form for the hadronic part \cite{dassinger2009complete}
\begin{equation}
\begin{split}
\Heff = {G_F \over \sqrt2}V_{cb} &\ H_\mu L^\mu + {\rm h.c}\cr
={G_F \over \sqrt2}V_{cb} & \biggl[(1+g_V)\cbar\gamma_\mu b + (-1+g_A)\cbar\gamma_\mu\gamma_5 b +  \dfrac{ g_S}{m_B}\ i \partial _\mu (\cbar b) + \dfrac{ g_P}{m_B}\ i \partial_\mu (\cbar\gamma_5 b) \biggr. \\
& \biggl.\quad  + \dfrac{g_T}{m_B}\ i\partial^\nu(\cbar i\sigma_{\mu\nu} b) + \dfrac{ g_{T5}}{m_B}\  i\partial^\nu(\cbar i\sigma_{\mu\nu}\gamma_5 b) \biggr]\  \ellbar\gamma^\mu(1-\gamma_5)\nu_\ell + {\rm h.c} \,.
\end{split}
\label{eq:Heff_NP}
\end{equation}

The terms added to our model will modify the helicity amplitudes used before. Using the current value of $|V_{cb}|$ given by the UTfit collaboration \cite{bona2006unitarity} \begin{equation}
|V_{cb} | = 0.04229 \pm 0.00057
\end{equation} we fit the data adding the additional factors $g_V$ and $g_A$ one at a time. For this fit we use the form factors used in the extrapolation of $|V_{cb}|$. We obtain a good fit with the values \begin{align}
g_V &= -0.01794 \pm 0.03926 \,, \\
g_A &= 0.05603 \pm 0.01838 \,.
\label{eq:NP_res}
\end{align} The values \ref{eq:NP_res} are obtained keeping the parameters of the CLN parametrization fixed. If we let them vary, the two parameters $g_V$ and $g_A$ don't contribute to the fit, and show a flat posterior likelihood. Thus, it is not possible to extract information from these data about the presence of deviations from the usual $V-A$ interaction. We have then decided to not pursue the search of new physics with tensorial operators on the same data.

\section{The $R_{D}$ and $R_{D^*}$ Anomalies}

The most statistically significant deviation involving B meson decays, at almost $4 \sigma $ deviation from the SM, is present in the two quantities $R_D$ and $R_{D^*}$, defined as \begin{equation}
R_{D^{(*)}} = \dfrac{\Gamma(\bar{B } \rightarrow D^{(*)} \tau \bar \nu_{\tau}  )}{\Gamma(\bar{B } \rightarrow D^{(*)} l \bar \nu_{l}  )}
\label{Eq:Ratio}
\end{equation} where $l$ is a light lepton: $l = e, \mu$. 
As reported by HFLAV collaboration in \cite{amhis2017averages}, current averages on these quantities are \begin{equation}
\begin{split}
R_{D} = 0.407 \pm 0.039 \pm 0.024 \,, \\
R_{D^*} = 0.304 \pm 0.013 \pm 0.007 \,.
\end{split}
\end{equation}

 The differential width for the process $B \rightarrow D^* \tau \nu_\tau$, can be split as \cite{bigi2017r} \begin{equation}
 \dfrac{d \Gamma_\tau}{dw} = \dfrac{d \Gamma_{\tau,1} }{dw} +  \dfrac{d \Gamma_{\tau,2} }{dw}
 \end{equation} where \begin{align}
 \dfrac{d \Gamma_{\tau,1} }{dw} = \left( 1 - \dfrac{m_\tau^2}{q^2} \right)^2 \left( 1 + \dfrac{m_\tau^2}{2 q^2} \right) \dfrac{d \Gamma}{dw} \,,\\
 \dfrac{d \Gamma_{\tau,2} }{dw} = k \sqrt{w^2 -1} \left( 1 - \dfrac{m_\tau^2}{q^2} \right)^2 \dfrac{3}{2} m_\tau^2 |H_t|^2 \,.
 \end{align}
   Here $d \Gamma/ dw$ is the same as in the extraction of $V_{cb}$, marginalized over the kinematical angles. It is then possible to split the ratio $R_{D^*}$ with this notation. Using the form factors obtained during the extraction of $|V_{cb}|$ with the BGL parametrization, we can obtain easily $R_{\tau,1}(D^*)$ with value \begin{equation}
 R_{\tau,1}(D^*) = 0.230 \pm 0.017 \; .
 \end{equation} This result is compatible with the one in \cite{bigi2017r}. Using the experimental value of $R^{exp}(D^*)$, we can estimate the value of $R_{\tau,2}$, obtaining \begin{equation}
 R_{\tau,2} = 0.080 \pm 0.028 \;.
 \label{eq:Rtau2}
 \end{equation} In order to calculate a theoretical prediction of this last observable, it is necessary to exploit the HQET ratio between form factors, in particular $R_0(w)$. In ref. \cite{fajfer2012b}  the value of the form factors ratio $R_0(1) = 1.14$ is used. The standard model prediction with this value is sensibly lower than the experimental result. We can try to fit the Wilson coefficient for a pseudo-scalar operator, since it can enhance the longitudinal contribution of the decay. 
 Adding it, the result we obtain  is \begin{equation}
 g_P( m_b ) = 9.058 \pm 4.666 \,.
 \end{equation} As previously pointed out, the ratios obtained in HQET are not completely reliable, they can have uncertainties which are difficult to estimate. We try then to let $R_0(1)$ vary around the value used before. In this trial we don't add the pseudo-scalar operator. We obtain a good fit with the value \begin{equation}
 R_0 (1) = 3.3923 \pm 0.343 \;.
 \end{equation}  The value obtained is roughly three times bigger than the one indicated by the HQET relations.
 \begin{figure}
 	\centering
 	\includegraphics[width=\textwidth]{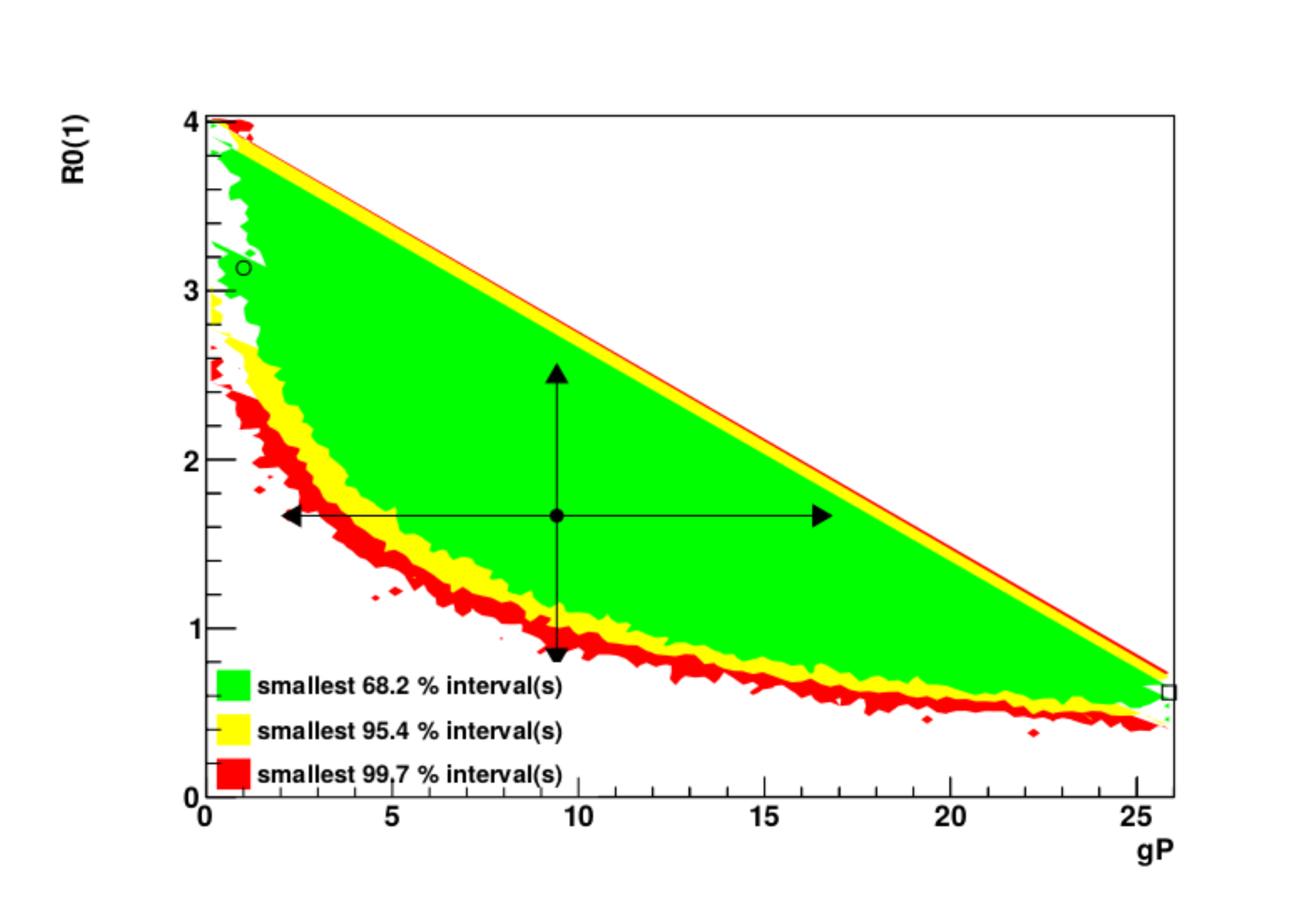}
 	\caption{Correlation diagram between the two parameter that can enhance the decay amplitude with a massive lepton, namely $g_P$ and $R_0(1)$.}
 	\label{fig:R01_gP}
 \end{figure}
We show then the correlation diagram between $g_P$ and $R_0(1)$, when are both included and free to vary, in figure \ref{fig:R01_gP}.

\section{Conclusions}

 The large discrepancy between the results of $|V_{cb}|$ from the CLN and BGL parametrization is a signal that the CLN expansions of $h_{A_1}$, $R_1$ and $R_2$ are too  constrained, without a reliable estimate of their uncertainties. With the improved precision of the recent data, and of the data that will come from the future run of Belle 2,  this procedure can't be reliably used for obtaining $V_{cb}$.
 
 We have explored the possibility of the presence of new physics at low energy, using an effective theory approach. While we found that it is not possible to extract information about  enhancements of the vectorial or axial channel  from  the data with massless leptons, we also obtained  a nonzero value for the pseudoscalar Wilson coefficient from the observable $R_{D^*}$. To estimate this contribution, HQET relations has been used. Surely, it is still necessary a better understanding of nonpertubative QCD to have a clear resolution of the anomaly.

\bibliographystyle{unsrt}

\bibliography{skeleton}

\end{document}